\documentclass[
    ,final            
  ]
  {aipproc}

\layoutstyle{8x11double}

\def\Gtsim{\mathrel{\vbox to 0pt{\hbox{$\sim$}}\llap{$>$}}}

\begin{document}

\title{Suzaku Observation of AXP 1E 1841-045 in SNR Kes 73}


\author{M.~Morii}{
  address={Rikkyo University, Nishi-ikebukuro 3-34-1, Toshima-ku, Tokyo 171-8501, Japan}
}

\author{S.~Kitamoto}{
  address={Rikkyo University, Nishi-ikebukuro 3-34-1, Toshima-ku, Tokyo 171-8501, Japan}
}

\author{N.~Shibazaki}{
  address={Rikkyo University, Nishi-ikebukuro 3-34-1, Toshima-ku, Tokyo 171-8501, Japan}
}

\author{D.~Takei}{
  address={Rikkyo University, Nishi-ikebukuro 3-34-1, Toshima-ku, Tokyo 171-8501, Japan}
}

\author{N.~Kawai}{
  address={Department of Physics, Tokyo Institute of Technology, Ookayama 2-12-1, Meguro-ku, Tokyo 152-8551, Japan}
}

\author{M.~Arimoto}{
  address={Department of Physics, Tokyo Institute of Technology, Ookayama 2-12-1, Meguro-ku, Tokyo 152-8551, Japan}
}

\author{M.~Ueno}{
  address={Department of Physics, Tokyo Institute of Technology, Ookayama 2-12-1, Meguro-ku, Tokyo 152-8551, Japan}
}

\author{Y.~Terada}{
  address={Cosmic Radiation Laboratory, Institute of Physical and Chemical Research (RIKEN), Wako, Saitama 351-0198, Japan}
}

\author{T.~Kohmura}{
  address={Physics Department, Kogakuin University 2665-1 Nakano-cho, Hachioji, Tokyo 192-0015, Japan}
}

\author{S.~Yamauchi}{
  address={Faculty of Humanities and Social Sciences, Iwate University, 3-18-34 Ueda, Morioka, Iwate 020-8550, Japan}
}


\begin{abstract}
Anomalous X-ray pulsars (AXPs) are thought to be magnetars,
which are neutron stars with ultra strong magnetic field of $10^{14}$--$10^{15}$ G. Their energy spectra below $\sim$10 keV are
modeled well by two components consisting of a blackbody (BB) ($\sim$0.4 keV) and rather steep power-law (POW) function
(photon index $\sim$2-4). Kuiper et al.(2004) discovered hard X-ray component above $\sim$10 keV from some AXPs.
Here, we present the Suzaku observation of the AXP 1E 1841-045 at the center of supernova remnant Kes 73.
By this observation, we could analyze the spectrum from
0.4 to 50 keV at the same time. Then, we could test whether the spectral model above was valid or not
in this wide energy range. We found that there were residual in the spectral fits when fit by the model of BB + POW.
Fits were improved by adding another BB or POW component. But the meaning of each component
became ambiguous in the phase-resolved spectroscopy.
Alternatively we found that NPEX model fit well for both phase-averaged spectrum
and phase-resolved spectra. In this case, the photon indices were constant during all phase,
and spectral variation seemed to be very clear. This fact suggests somewhat fundamental meaning for
the emission from magnetars.
\end{abstract}


\maketitle


\section{Introduction}
Anomalous X-ray pulsars (AXPs) are thought to be magnetars,
which are strongly magnetized ($\sim 10^{14}$ -- $10^{15}$ G)
neutron stars with emissions
powered by the dissipation of the magnetic energy
(see \cite{Woods Thompson 2006} for a review).
The spectra of the AXP had been modelled well by 
the compound of a blackbody and a power-law function
below $\sim 10$ keV region.
Besides, two temperature blackbody spectrum was also fit well for some AXPs
\cite{Morii et al 2003, Naik et al 2007}.
On the other hand, separate hard X-ray emission was discovered for some AXPs above $\sim 10$ keV
\cite{Kuiper Hermsen Mendez 2004, Kuiper et al 2006}.

AXP 1E 1841-045 is located on the center of the supernova remnant (SNR) Kes 73
(G 27.4+0.0) with diameter of about 4$^\prime$.
The kinematic distance toward the SNR was estimated to be between
6 and 7.5 kpc \cite{Sanbonmatsu Helfand 1992}.
The pulse period of 1E 1841-045 was 11.8 s \cite{Vasisht Gotthelf 1997}.
Morii et al. (2003) \cite{Morii et al 2003} reported that the spectrum was fitted well
with the model consisting of the blackbody ($kT = 0.44 \pm 0.002$ keV)
and the power-law function with the hardest photon index among
AXP ($\Gamma = 2.0 \pm 0.3$), using Chandra data (0.6 -- 7.0 keV).
In this analysis, they also showed two blackbody model fits well.
Kuiper, Hermsen, \& Mendez (2004) \cite{Kuiper Hermsen Mendez 2004}
discovered hard X-ray emission
up to $\sim 150$ keV by using RXTE.

\section{Observation}
1E 1841$-$045 and Kes 73 were observed
by Suzaku on 2006 April 19--22,
as a target of AO-1 (PI and Co-PI were I.~Harrus and M.~Morii).
Suzaku \cite{Mitsuda et al 2007} has two types of X-ray detectors in operation:
the X-ray Imaging Spectrometer (XIS; \cite{Koyama et al 2007}) and
the Hard X-ray Detector (HXD; \cite{Takahashi et al 2007, Kokubun et al 2007}).
We chose the ``1/8 window mode'' for XIS
to obtain the timing resolution of 1 s, which corresponds to 0.08 period
interval of this AXP's pulsation.
HXD is composed of the PIN photodiode and GSO scintillators
mounted in the well of the collimator shield,
whose parts cover 10$-$70 keV and 40$-$600 keV, respectively.


This observation was pointed at
(R.A., Dec.) $ = (18^{\rm h} 41^{\rm m} 15.5^{\rm s}, -4^\circ 51^{\prime} 24.5^{\prime\prime})$
in the HXD nominal pointing mode.
Total exposure time requested was 100 ks.

\section{Analysis}


We used cleaned events made by the standard pipeline processing of rev-1.2.
We selected the source and background regions as the concentric circle 
and annulus whose central positions were at the peak position of the AXP.
The source circular radius was set to be 4.33$^\prime$,
and the radii of the inner and outer boundaries of
the background annulus were set to be 4.33$^\prime$ and 6.00$^\prime$.
For the PIN analysis, we used background file made by LCFIT (bgd\_d) method of version 1.2ver0611.
We also applied the additional GTI (good time interval) selection
as well as ordinary GTI made by event and background files.
The net exposures for XISs and PIN were 95.3 ks and 57.8 ks.

We checked light curves of all XISs and PIN binned by 300, 10 and 1 s.
We found no special characteristics due to background variation, instrumental troubles 
and activities in the AXP.


\section{Timing analysis}

\begin{figure}
    \includegraphics[width=55mm, angle=270]
        {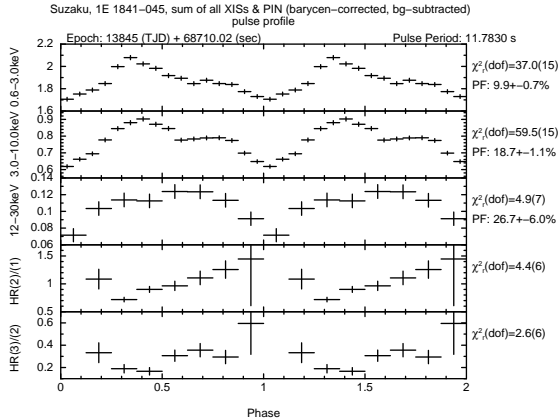}
  \caption{Pulse profiles in the energy ranges of
  $0.6 - 3.0$ keV, $3.0 - 10.0$ keV, and  $12.0 - 30.0$ keV were shown
  in the 1st, 2nd and 3rd panels. Pulse profiles of the 1st and
  2nd band were double peaked, while the 3rd was single peaked.
  The 4th and 5th panels show the hardness ratios of 2nd/1st and 3rd/2nd,
  respectively.
}\label{fig: prof+hr.ps}
\end{figure}

We corrected the photon arrival times into those at the solar barycenter by
using \texttt{aebarycen} (ver. 2006-08-02).
We searched periodicity for this light curve by using the \texttt{xronos/powspec},
and we determined the pulse period precisely by using the \texttt{xronos/efsearch}.
The obtained pulse period was 11.7830(2) s
at the epoch of 13845 (TJD) plus 68704.31 s.
%
We made the pulse profiles for some energy ranges
(0.6 -- 3.0 keV, 3.0 -- 10.0 keV, and 12.0 -- 30.0 keV)
(Figure \ref{fig: prof+hr.ps}).


%

\section{Energy Spectrum}

\subsection{Phase-averaged Spectroscopy}

Due to point spread function (PSF) of XRTs (HPD = 2$^\prime$)
\cite{Serlemitsos et al 2007},
AXP's spectrum could not be spatially separated from
SNR Kes 73. Then at first we determined the SNR
spectral model by using Chandra archival data.
It was modeled well by a VSEDOV model. We used this model as the SNR
model in the followings.
We then fit the spectrum by
$c_{\rm det} \times \left( c_{\rm SNR} \times {\rm SNR} + {\rm AXP} + {\rm CXB} + {\rm GRXE} \right)$.

where $c_{\rm det}$ was normalization of detectors. $c_{\rm SNR}$ was normalization of SNR.
CXB and GRXE mean components from the cosmic X-ray background and galactic ridge X-ray emission.
CXB + GRXE was taken into consideration only for HXD/PIN. CXB + GRXE was estimated by using another Suzaku data
of GRXE observation near this field ($l = 28.5^\circ$, $b = -0.2^\circ$).
For AXP component, we tried to fit by a power-law function (POW),
a blackbody (BB), and BB + BB. These models were denied.
BB + POW model fit better, but wavy residuals remained.
Of cause, addition of another component like POW or BB could improve the fit.
But for BB + POW + BB model, parameters were physically unacceptable.
BB + POW + POW showed possible solution(Table \ref{tab: a}).

Alternatively, we noticed that the NPEX model could give good solution (Table \ref{tab: a}).
NPEX is following function:
\begin{equation}
(A_n E^{-\alpha} + A_p E^{+\beta}) \exp ( - \frac{E}{kT}).
\end{equation}
This model was originally developed to explain the spectra of accretion-powered pulsars
by Makishima et al. (1999) \cite{Makishima et al 1999}.
It approximates spectrum for an unsaturated thermal Comptonization. 

\begin{table}
\begin{tabular}{llllll}
\hline
    \tablehead{1}{l}{b}{MODEL}
  & \tablehead{1}{l}{b}{$n_H$}
  & \tablehead{1}{l}{b}{kT}
  & \tablehead{1}{l}{b}{$R_{\bf BB}$}
  & \tablehead{1}{l}{b}{photon index}
  & \tablehead{1}{l}{b}{$\chi^2$/dof} \\

  & \tablehead{1}{l}{b}{($\times 10^{22}$ cm$^{-2}$)}
  & \tablehead{1}{l}{b}{(keV)}
  & \tablehead{1}{l}{b}{(km) @ 7kpc}
  &
  &  \\
\hline
BB + POW & $2.70\pm 0.01$ & $0.370\pm0.006$ & $8.4\pm 0.3$ & $2.24 \pm 0.04$ & $1616.7/902 = 1.79$ \\
BB + BB + POW & $2.84\pm 0.01$ & $0.15$ & $89 \pm 4$ & $3.30 \pm 0.03$ & $1354.6/900 = 1.51$ \\
              &                & $4.9 \pm 0.2$ & $0.041 \pm 0.002$ &  \\
BB + POW + POW & $2.87\pm 0.02$ & $0.54 \pm 0.02$ & $2.5 \pm 0.3$ & $5.0 \pm 0.3$ & $1312.7/900 = 1.46$ \\
              &                 &                 &               & $1.6 \pm 0.1$ \\ \hline
    \tablehead{1}{l}{b}{MODEL}
  & 
  & \tablehead{1}{l}{b}{Folding energy}
  & \tablehead{1}{l}{b}{photon index}
  & \tablehead{1}{l}{b}{photon index}
  & \\ 

  & 
  & \tablehead{1}{l}{b}{(keV)}
  & \tablehead{1}{l}{b}{($\alpha$)}
  & \tablehead{1}{l}{b}{($-\beta$)}
  & \\ \hline
NPEX & $2.82\pm 0.01$ & $58 \pm 1$ & $3.54\pm 0.04$ & $0.92 \pm 0.06$ & $1352.2/901 = 1.50$ \\
\hline
\end{tabular}
\caption{Results of phase-averaged spectral fits}
\label{tab: a}
\end{table}


\begin{figure}
    \includegraphics[width=55mm, angle=270]
        {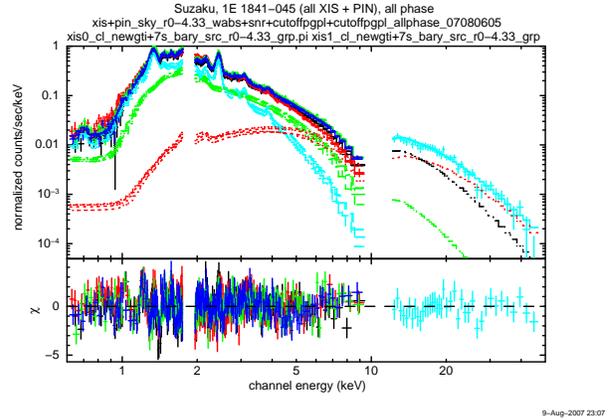}
  \caption{Phase-averaged spectrum fit by NPEX model as AXP component.}
  \label{fig: xis+pin_sky_r0-4.33_wabs+snr+cutoffpgpl+cutoffpgpl_allphase_07080605.ps}	
\end{figure}

\begin{figure}
    \includegraphics[width=55mm, angle=270]
        {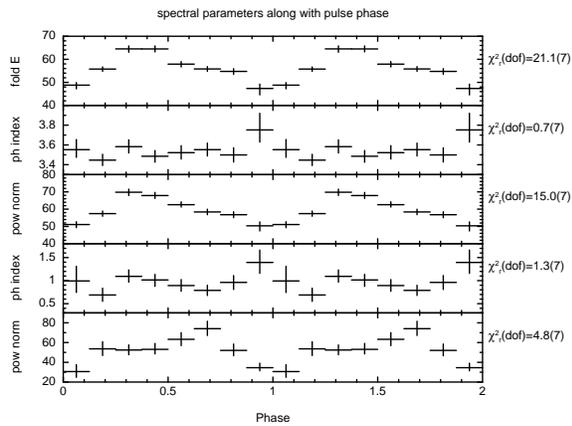}
  \caption{Variation of spectral parameter along with pulse phase when fit with NPEX model.}
\label{fig: specvar_phase_1sigma_07080605.ps}
\end{figure}

\subsection{Pulse-phase resolved spectroscopy}

When fit by BB+POW or BB+POW+POW, the variation of parameters in phases was not clear.
All parameters varied. On the other hand, when fit by NPEX model, photon indices were
consistent with constant (Figure \ref{fig: specvar_phase_1sigma_07080605.ps}, 2nd and 4th panels),
while normalizations of power-law
(Figure \ref{fig: specvar_phase_1sigma_07080605.ps}, 3rd and 5th panels)
and folding energy (Figure \ref{fig: specvar_phase_1sigma_07080605.ps}, 1st panel) varied significantly.
The first peak was dominant by the soft power-law component,
while the second peak was dominant by the hard one. This result was consistent with pulse profile
(Figure \ref{fig: prof+hr.ps}).
The folding energy also varied in the same way as the pulse profile.

\section{Discussion}

It is interesting to compare the NPEX parameters between this AXP and accretion-powered pulsars.
$\alpha$ and $-\beta$ were $\sim0$ and $\sim -2$ for accretion-powered pulsars,
while those are $3.54 \pm 0.04$ and $0.92 \pm 0.06 $ for this AXP (Table \ref{tab: a}).
Both of this AXP are larger than those of accretion powered pulsars by about three.
For accretion powered pulsars,
power-law component with $\alpha \sim 0$ supplies seed photons for Comptonization.
On the other hand, for this AXP,
the soft power-law component with $\alpha = 3.54 \pm 0.04$ would supply seed photons.
This steep power-law may be a modified blackbody,
which is originated from magnetar's surface and deformed
by strong magnetic field of magnetar \cite{Ozel 2001}.
The $\beta$ parameters can be interpreted as follows.
The Comptonization is parameterized by $y$ parameter \cite{Rybicki Lightman 1979}.
For $y >> 1$, $\beta$ becomes 2,
meaning that the spectrum becomes saturated.
For accretion-powered pulsars,
the emission comes from the accretion column where the plasma density is large,
so $y$ become large.
On the other hand, the plasma density of this AXP is smaller
than those of accretion-powered pulsars.
The $y$ parameter become small, then $\beta$ becomes small.
The $y$ must be $\Gtsim 1$ in this AXP, so that Comptonization process can be effective.
This fact suggests that there are corona around magnetosphere of this magnetar
\cite{Beloborodov Thompson 2007}.

\bibliographystyle{aipproc}   




\end{document}